\documentclass[12pt,english]{article}
\pdfoutput=1
\usepackage[T1]{fontenc}
\usepackage{graphicx,array}
\usepackage{cite}
\usepackage{color}
\usepackage{bm}
\usepackage{dsfont}
\usepackage{latexsym}
\usepackage{amsthm}
\usepackage{amsmath}
\usepackage{stackengine}
\usepackage[titletoc]{appendix}
\usepackage{enumitem}
\usepackage{amssymb}
\usepackage{float}
\usepackage[unicode=true,
 bookmarks=true,bookmarksnumbered=false,bookmarksopen=false,
 breaklinks=false,pdfborder={0 0 1},backref=false,colorlinks=true]
 {hyperref}
\hypersetup{
 pdfauthor={Avik Banerjee, Achilleas P. Porfyriadis, Grant N. Remmen},
 citecolor=blue,linkcolor=black,urlcolor=black}
\usepackage{breakurl}
\usepackage[hang,flushmargin]{footmisc}

\newcolumntype{L}[1]{>{\raggedright\let\newline\\\arraybackslash\hspace{0pt}}m{#1}}
\newcolumntype{C}[1]{>{\centering\let\newline\\\arraybackslash\hspace{0pt}}m{#1}}

\def\simgt{\mathrel{\lower2.5pt\vbox{\lineskip=0pt\baselineskip=0pt
           \hbox{$>$}\hbox{$\sim$}}}}
\def\simlt{\mathrel{\lower2.5pt\vbox{\lineskip=0pt\baselineskip=0pt
           \hbox{$<$}\hbox{$\sim$}}}}

\newcommand{\be}{\begin{equation}}
\newcommand{\ee}{\end{equation}}
\newcommand{\bea}{\begin{eqnarray}}
\newcommand{\eea}{\end{eqnarray}}

\newcommand*\oline[1]{%
  \vbox{%
    \hrule height 0.5pt
    \kern0.68ex
    \hbox{%
      \kern-0.1em
      \ifmmode#1\else\ensuremath{#1}\fi
      \kern-0.1em
    }
  }
}

\definecolor{nicered}{rgb}{0.7,0.1,0.1}
\definecolor{nicegreen}{rgb}{0.1,0.5,0.1}
\hypersetup{
 colorlinks,citecolor=black,linkcolor=black,urlcolor=black}
\usepackage{xcolor}

\begin{document}

\interfootnotelinepenalty=10000
\baselineskip=18pt
\hfill CCTP-2024-17 

\hfill ITCP-IPP 2024/17

\vspace{2cm}
\thispagestyle{empty}
\begin{center}
{\LARGE \bf
Accidental Symmetry Near\\[1mm] Extreme Spinning Black Holes
}\\
\bigskip\vspace{1cm}{
{\large Avik Banerjee,${}^a$ Achilleas P. Porfyriadis,${}^{a}$ and Grant N. Remmen${}^{b}$}
} \\[7mm]
{
\it ${}^a$Crete Center for Theoretical Physics, Institute of Theoretical and Computational Physics,
Department of Physics, University of Crete, 70013 Heraklion, Greece\\[1.5 mm]
${}^b$Center for Cosmology and Particle Physics, Department of Physics,\\ New York University, New York, NY 10003, USA}
 \end{center}

\bigskip
\centerline{\large\bf Abstract}

\begin{quote}

Near the horizon of extremal charged black holes, an accidental symmetry is known to act on zero-temperature perturbations, transforming them into finite-temperature ones.
In this paper, we uncover the corresponding accidental symmetry 
of the vacuum Einstein equation 
near the horizon of extremal spinning black holes.
To do so, we first devise a new method of deriving the symmetry near the ${\rm AdS}_2 \times S^2$ near-horizon geometry of extreme Reissner-Nordstr\"om, using a new scaling coordinate transformation that unifies the near-horizon limits of extremal and near-extremal black holes in a way that is regular at zero temperature.
We use our new method to obtain the accidental symmetry in the near-horizon of extreme Kerr (NHEK).
We show that accidental symmetries combine neatly with the near-horizon isometries inside a Virasoro algebra.

\end{quote}
	
\setcounter{footnote}{0}

\newpage
\tableofcontents

\section{Introduction}

The near-horizon region of extreme and near-extreme Reissner-Nordstr\"om (RN) black holes, which has the geometry of ${\rm AdS}_2\times {\rm S}^2$, describes a universe of its own given by an exact solution of the Einstein-Maxwell equations that is often called the Bertotti-Robinson (BR) universe~\cite{Carter:2009nex}. 
In this universe, clever physicists with a good understanding of general relativity would quickly infer the existence of solutions with different asymptotics from their own ${\rm AdS}_2$ ones. They would also reckon that such solutions must belong to a phase space described by more parameters than the one setting the size of their universe. 
Specifically, this could be done as follows. First, the BR physicists would compute the spherically symmetric gravitational perturbations of their universe and find a four-parameter family of solutions with all four parameters entering the perturbation of the size of the $S^2$ transverse to ${\rm AdS}_2$. Then, following Ref.~\cite{Hadar:2020kry}, they would classify these solutions into a singlet and a triplet of the ${\rm SL}(2)$ symmetry group of ${\rm AdS}_2$. Only the singlet solution has a back reaction that is consistent with the ${\rm AdS}_2$ asymptotics, corresponding to another BR universe of different size~\cite{Maldacena:1998uz}. The ${\rm SL}(2)$ triplet, on the other hand, does not respect the  ${\rm AdS}_2$ asymptotics and upon back reaction changes them to the flat asymptotics of RN black holes. This phenomenon was studied in Ref.~\cite{Hadar:2020kry}, where it was called an anabasis off BR.\footnote{In the context of two-dimensional models of gravity, such as the Jackiw-Teitelboim  theory\cite{Jackiw:1984je, Teitelboim:1983ux}, anabasis manifests itself as a modification of the ${\rm AdS}_2$ asymptotics to those of so-called nearly-${\rm AdS}_2$ spacetimes\cite{Almheiri:2014cka, Jensen:2016pah, Maldacena:2016upp, Engelsoy:2016xyb}; see Refs.~\cite{Mertens:2022irh, Sarosi:2017ykf} for reviews.} The invariant ${\rm SL}(2)$ Casimir associated with any particular triplet solution may be used as a local diagnostic that differentiates between the extreme and near-extreme RN: if the Casimir vanishes, back reaction leads to an extreme RN (ERN) solution, whereas if it is positive, back reaction builds a near-extreme RN geometry.

In Ref.~\cite{Porfyriadis:2021psx}, it was shown that there exists an accidental symmetry that maps BR perturbations with vanishing ${\rm SL}(2)$ Casimir onto perturbations with positive Casimir. These accidental symmetries, which act as linearized diffeomorphisms on the perturbative solutions, may be thought of as on-shell large diffeomorsphisms (asymptotic symmetries) of ${\rm AdS}_2$. They are large because they map one physical state to another: perturbations that, upon back reaction, give rise to extreme black holes are mapped onto perturbations corresponding to near-extreme black holes. They are on-shell because they ensure that the Einstein equation is satisfied beyond the ${\rm AdS}_2$ region as well, across the boundary of ${\rm AdS}_2$, allowing for smooth gluing onto the asymptotically flat region of RN black holes.

In this paper, we find the analogous accidental symmetry in the gravitating box that is the near-horizon of extreme Kerr (NHEK)~\cite{Bardeen:1999px}, an exact solution of the vacuum Einstein equation with back reaction properties similar to BR~\cite{Amsel:2009ev, Dias:2009ex}. Here too, axisymmetric gravitational perturbations of NHEK can lead to anabasis towards an extreme Kerr (EK) or a near-extreme Kerr.
A direct search for the accidental symmetries of NHEK, similar to the one performed in Ref.~\cite{Porfyriadis:2021psx} for BR, is significantly complicated by the following two aspects. 
First, identifying a simple local diagnostic for the extremality condition for a given anabasis perturbation is less straightforward because, unlike from the BR case, none of the metric perturbation components for NHEK is gauge invariant.
More importantly, solving directly for the accidental symmetry by plugging into the Einstein equation, as it was done in Ref.~\cite{Porfyriadis:2021psx}, is substantially more challenging for a rotating spacetime such as NHEK because separation of the angular variables is difficult to ensure in metric formalism.
Therefore, in this paper we devise a new method for identifying the accidental symmetry of RN, detailed in Sec.~\ref{sec:RN}, which is then straightforwardly generalized to the case of Kerr in Sec.~\ref{sec:Kerr}. The method emphasizes constructing a one-parameter family of scaling coordinates, used for taking the near-horizon near-extremality limit from Kerr to NHEK, which are such that the coordinate transformation from Boyer-Lindquist Kerr coordinates to Poincar\'e NHEK coordinates is regular in the limit of vanishing Hawking temperature. This allows us to both readily distinguish extreme from near-extreme anabasis perturbations off NHEK as well as directly derive the desired accidental symmetry of NHEK. 
Finally, we clarify the fact that the precise form of the vector fields generating the accidental symmetry depends on the gauge in which they are acting. In particular, this fact allows us to pick a gauge for the extreme anabasis solutions such that the accidental symmetry acting on them to produce the near-extreme anabasis solution naturally takes the form of the $L_{-2}$ mode in the Virasoro algebra.
This paper opens several promising avenues for future research into the near-horizon dynamics of near-extremal black holes, discussed in Sec.~\ref{sec:discussion}.

\section{Accidental symmetry in the near-horizon of extreme RN}\label{sec:RN}

In Ref.~\cite{Porfyriadis:2021psx}, the accidental symmetries in the near-horizon of RN black holes of extremal charge were found by searching directly for maps between solutions of the linearized Einstein equation around BR. In this section, we present a different derivation of the electrovacuum accidental symmetry associated with the near-horizon of (near-)extreme RN black holes. The main advantage of the alternative derivation here is that it is amenable to generalization to the case of Kerr black holes.

We begin with an expansion of ERN around BR,
\begin{align}\label{ERN series}
	g^{\rm ERN}=g^{\rm BR}+\lambda \, h +\mathcal{O}(\lambda^2).
\end{align}
Here $\lambda$ is a scaling parameter used to write the ERN spacetime in a one-parameter family of coordinates that facilitate taking the near-horizon limit of ERN via the scaling limit $\lambda\to 0$.
This ensures that $h$ is a solution to the linear Einstein equation around BR, $\mathcal{E}(g^{\rm BR},h)=0$. In the language of Ref.~\cite{Hadar:2020kry}, $h$ is an anabasis solution off BR whose back reaction builds the exterior asymptotically flat region of ERN.  
Next, consider a linearized diffeomorphism generated by a vector field $\xi$.
Ref.~\cite{Porfyriadis:2021psx} asked whether its action on $h$ can ever produce a physically distinct solution of the same linear Einstein equation around BR, that is to say, for which $\xi$ we have $\mathcal{E}(g^{\rm BR},\mathcal{L}_\xi h)=0$. General covariance of the Einstein equation implies that one expects to find as solutions any vectors $\xi$ that are isometries of the background $g^{\rm BR}$, such as the ${\rm SL}(2)$ symmetry generators,
\begin{align}\label{SL2 vectors}
\xi_{1}=\partial_t\,, \quad 
\xi_0=t\partial_t-r\partial_r\,, \quad
\xi_{-1}=\left(t^2+\frac{1}{r^2}\right)\partial_t-2rt\partial_r,
\end{align}
with algebra
\begin{align}\label{SL2 algebra}
	[\xi_{\pm 1}, \xi_0]=\pm \xi_\pm\,,\qquad [\xi_1,\xi_{-1}]=2\xi_0.
\end{align}
However, even within spherical symmetry in the electrovacuum equations, it turns out that there exists an additional solution, $\xi_{-2}$, that produces a physically distinct perturbation. This accidental symmetry of the linear Einstein equation around BR may be understood as follows. Let $\mu$ be the size of the linearized diffeomorphisms generated by the $\xi$ vectors. We have that $\xi_{-2}$ produces the solution $h\to h+\mu\,\mathcal{L}_{\xi_{-2}}h$ that is an anabasis solution off BR whose back reaction builds the exterior asymptotically flat region of near-extreme RN with charge-to-mass ratio controlled by $\mu$. In other words, we have that near-extreme RN may be written as a double series expansion in $\lambda$ and $\mu$ as follows:
\begin{align}\label{RN double series}
	g^{\rm RN}=g^{\rm BR}+\lambda\left(h+\mu\,\mathcal{L}_{\xi_{-2}}h\right)+\cdots.
\end{align}
In Ref.~\cite{Porfyriadis:2021psx}, $\xi_{-2}$ was found by solving the linear Einstein equation $\mathcal{E}(g^{\rm BR},\mathcal{L}_\xi h)=0$. Here we will obtain $\xi_{-2}$ by finding an appropriate one-parameter family of coordinates, which will allow us to write an RN spacetime of charge-to-mass ratio $\sqrt{1-\lambda^2\kappa^2}$, with $\mu=4\kappa^2$, as a double series of the form in Eq.~\eqref{RN double series}.

Comparing Eqs.~\eqref{ERN series} and \eqref{RN double series}, it is evident that the existence of the accidental symmetry of the linear Einstein equation around BR is due to the fact that both extreme and near-extreme RN black holes have the same BR solution as appropriate near-horizon limits. Indeed, one does not necessarily need in Eq.~\eqref{RN double series} the series in $\mu$ in order to produce BR in the $\lambda\to 0$ limit. Given a charge-to-mass ratio of $\sqrt{1-\lambda^2\kappa^2}$, the near-extremality limit is ensured by the series in $\lambda$ and one can obtain, for finite $\kappa$, an expansion for near-extreme RN around BR that is directly analogous to Eq.~\eqref{ERN series}: $g^{\rm RN}=\tilde{g}^{BR}+\lambda \, \tilde{h} +\mathcal{O}(\lambda^2)$\cite{Spradlin:1999bn}. However, this yields BR in a set of coordinates that are not continuously connected to the ones in Eq.~\eqref{RN double series} in the sense that the coordinate transformation between $\tilde{g}^{BR}$ and $g^{\rm BR}$, which may be found in Ref.~\cite{Hadar:2020kry}, is not regular in the $\kappa\to 0$ limit. Indeed, $g^{\rm BR}$ and $\tilde{g}^{BR}$ in the standard treatment, such as the one of Ref.~\cite{Hadar:2020kry}, present in Poincar\'e and Rindler coordinates, respectively. The transformation between them is then singular in the limit of vanishing Rindler temperature.

\subsection{Derivation}

Consider the magnetic RN black hole solution, 
\begin{align}
	\begin{aligned}
	{\rm d}s^2&=-\left(1-\frac{2M}{\hat{r}}+\frac{Q^2}{\hat{r}^2}\right){\rm d}\hat{t}^2+
	\left(1-\frac{2M}{\hat{r}}+\frac{Q^2}{\hat{r}^2}\right)^{-1}{\rm d}\hat{r}^2+
	\hat{r}^2 {\rm d}\Omega^2 \\
	F&=Q\sin\theta \, {\rm d}\theta\wedge {\rm d}\phi.
	\end{aligned}
\end{align}
For ERN, defined by $Q=M$, the series \eqref{ERN series} may be obtained by using the following one-parameter family of coordinates (in $M=1$ units),
\begin{align}\label{ERN coord map}
	\hat{t}=t/\lambda+\mathcal{O}(1), \quad \hat{r}=1+\lambda r+\mathcal{O}(\lambda^2).
\end{align}
This produces the first two series terms shown in Eq.~\eqref{ERN series} with $g^{\rm BR}$ given by AdS$_2$ in Poincar\'e coordinates,
\begin{align}\label{BR Poincare}
	g^{\rm BR}_{\mu\nu}{\rm d}x^\mu {\rm d}x^\nu=-r^2 {\rm d}t^2+\frac{{\rm d}r^2}{r^2}+{\rm d}\Omega^2,
\end{align}
and $h$ given by
\begin{align}\label{h100}
	h_{\mu\nu}{\rm d}x^\mu {\rm d}x^\nu=2r^3 {\rm d}t^2+\frac{2}{r}{\rm d}r^2+2r \, {\rm d}\Omega^2.
\end{align}
In the above, $h_{\theta\theta}$ is gauge invariant as a perturbation of $g^{\rm BR}_{\theta\theta}$, whereas $h_{tt},h_{tr},h_{rr}$ are not. They may be adjusted by appropriately choosing the terms subleading in $\lambda$ in Eq.~\eqref{ERN coord map}. Indeed, the most general spherically symmetric pure gauge shift of Eq.~\eqref{h100} by $\mathcal{L}_\zeta g^{\rm BR}$ may be obtained by replacing Eq.~\eqref{ERN coord map} with
\begin{align}\label{ERN general coord map}
	\hat{t}=t/\lambda+\zeta^t(t,r)+\mathcal{O}(\lambda), \quad \hat{r}=1+\lambda r+\lambda^2 \zeta^r(t,r)+\mathcal{O}(\lambda^3),
\end{align}
so that Eq.~\eqref{h100} becomes
\begin{align}\label{h100-general}
	\begin{aligned}
	&h_{tt}=2r^3-2r\left(\zeta^r+r\partial_t\zeta^t\right)\\
	&h_{tr}=-r^2\partial_r\zeta^t+\frac{1}{r^2}\partial_t\zeta^r\\
	&h_{rr}=\frac{2}{r}-\frac{2}{r^3}\left(\zeta^r-r\partial_r\zeta^r\right)\\
	&h_{\phi\phi}=\sin^2\theta \, h_{\theta\theta}=2r \sin^2\theta.
	\end{aligned}
\end{align}
This perturbation solves the linear Einstein equation $\mathcal{E}(g^{\rm BR},h)=0$ for arbitrary $(\zeta^t,\zeta^r)$.

Now let us consider a near-extreme RN with charge-to-mass ratio
\begin{align}
	{Q\over M}=\sqrt{1-\lambda^2\kappa^2},
\end{align}
 and follow up the transformation in Eq.~\eqref{ERN coord map} by
\begin{align}\label{xi diffeo}
	t\to t+\kappa^2\xi^t(t,r), \quad 
	r\to r+\kappa^2\xi^r(t,r).
\end{align}
That is to say, write RN in the one-parameter family of coordinates 
\begin{align}\label{RN nice coords}
\begin{aligned}
		&\hat{t}={t\over\lambda}+{\kappa^2\over\lambda}\xi^t(t,r)\\ &\hat{r}=1+\lambda r+\lambda\kappa^2 \xi^r(t,r),
\end{aligned}
\end{align}
and expand in both $\lambda$ and $\kappa$,
\begin{align}\label{RN expansion scheme}
	g^{\rm RN}=g^{\rm BR}+\kappa^2\,G +\lambda h +\lambda\kappa^2\,H+\cdots.
\end{align}
Then $g^{\rm BR}$ and $h$ are given as in Eqs.~\eqref{BR Poincare} and \eqref{h100}, respectively, while $G$ is given by
\begin{align}\label{G}
	\begin{aligned}
	G_{tt}&=1-2r(\xi^r+r\partial_t\xi^t) \\
	G_{tr}&=-r^2 \partial_r\xi^t +\frac{1}{r^2}\partial_t\xi^r\\
	G_{rr}&=\frac{1}{r^4}-\frac{2}{r^3}\left(\xi^r-r\partial_r\xi^r\right).
	\end{aligned}
\end{align}
Notice that if we set $\lambda=0$ in Eq.~\eqref{RN expansion scheme}, then we get a linearization of the exact solution on the left-hand side around $g^{\rm BR}$, and therefore $G$ solves the linearized Einstein equation $\mathcal{E}(g^{\rm BR}, G)=0$ for arbitrary $\xi$. Similarly, of course, setting $\kappa=0$ implies $\mathcal{E}(g^{\rm BR}, h)=0$, which we already knew from Eq.~\eqref{ERN series}.
On the other hand, $H$, which also depends on $\xi$, is not in general a linear solution around $g^{\rm BR}$. However, we can ensure that it is a solution, $\mathcal{E}(g^{\rm BR}, H)=0$, if we choose $\xi$ such that $G$ vanishes,
\begin{align}\label{xi-2 with SL2 added}
	G=0 \,\,\Rightarrow\,\,
	\xi=\frac{1}{12}\left(t^3+\frac{3t}{r^2}\right)\partial_t
	-\frac{1}{4}r\left(t^2-\frac{1}{r^2}\right)\partial_r
	+e_0\xi_0+e_1\xi_{1}+e_{-1}\xi_{-1}.
\end{align}
For such $\xi$ we obtain
\begin{align}\label{RN series naive gauge}
	g^{\rm RN}=g^{\rm BR}+\lambda \left(h +\kappa^2\,H\right)+\cdots,
\end{align}
which implies that $h+\kappa^2 H$ is a linear solution around $g^{\rm BR}$, and since so is $h$, we have that $H$ must be a solution too.
That is to say, the most general $\xi$ that allows us to use our one-parameter family of coordinates \eqref{RN nice coords} in order to write the RN solution in the double series form in Eq.~\eqref{RN series naive gauge} is given by
\begin{align}\label{xi-2}
	\Xi=\frac{1}{12}\,\xi_{-2}  \quad\textrm{with}\quad \xi_{-2}=\left(t^3+\frac{3t}{r^2}\right)\partial_t
	-3r\left(t^2-\frac{1}{r^2}\right)\partial_r,
\end{align}
together with the ${\rm SL}(2)$ generators \eqref{SL2 vectors}. Specifically, for Eq.~\eqref{xi-2} we have
\begin{align}\label{H naive}
	H_{\mu\nu}{\rm d}x^\mu {\rm d}x^\nu=
	-\frac{r^3}{2}\left(t^2-\frac{1}{r^2}\right) {\rm d}t^2 -4t {\rm d}t {\rm d}r
	-\frac{1}{2r}\left(t^2-\frac{1}{r^2}\right) {\rm d}r^2
	-\frac{r}{2}\left(t^2-\frac{1}{r^2}\right) \, {\rm d}\Omega^2.
\end{align}
We see that Eq.~\eqref{RN series naive gauge} has very similar features to Eq.~\eqref{RN double series}. First, $h$ is an anabasis solution off BR whose back reaction, if left alone, would build the asymptotically flat ERN. Next, $h+\kappa^2H$ is an anabasis solution off BR whose back reaction builds the near-extreme RN. Indeed, recall that the invariant ${\rm SL}(2)$ Casimir $\mu$, associated with any linear solution around BR whose $\theta\theta$ metric perturbation reads $ar+brt+cr(t^2-1/r^2)$, is given by $\mu=b^2-4ac$ \cite{Hadar:2020kry}, so that for Eq.~\eqref{h100} we have $h_{\theta\theta}=2r$ with $\mu=0$, while for Eq.~\eqref{H naive} we have  $h_{\theta\theta}+\kappa^2H_{\theta\theta}= 2r-\frac{\kappa^2}{2}r\left(t^2-\frac{1}{r^2}\right)$ with $\mu=4\kappa^2$.
Finally, notice that the derivation of Eq.~\eqref{RN series naive gauge} hinges on the existence of the accidental solution $\Xi$ for the equations $G=0$, which are the ones responsible for rendering $H$ a linear solution around BR.\footnote{Note that had we used the most general solution \eqref{xi-2 with SL2 added} in place of Eq.~\eqref{xi-2} to produce a more general $H$ than  in Eq.~\eqref{H naive}, the ${\rm SL}(2)$ invariant Casimir $\mu$ associated with $h+\kappa^2H$ would indeed remain invariant to leading order in $\kappa^2$.}

\subsection{Aligning the gauge}

The only difference between Eqs.~\eqref{RN series naive gauge} and \eqref{RN double series} is that the pair of solutions in Eqs.~ \eqref{h100} and \eqref{H naive} do not satisfy the relation $H=\mathcal{L}_{\Xi}h$ via Eq.~\eqref{xi-2}. This relation, which was the starting point in Ref.~\cite{Porfyriadis:2021psx}, is essential for interpreting $\Xi$ as an accidental symmetry because it implies that $\Xi$ acts as a linearized diffeomorphism that maps one solution $h$ onto a different solution $H$. 
Indeed, the $\xi_{-2}$ in Eq.~\eqref{xi-2} differs slightly from the one found in Ref.~\cite{Porfyriadis:2021psx}: in $\xi^t$ the term subleading in $1/r$ carries a different numerical factor.
Nevertheless, our $\xi_{-2}$ here, in Eq.~\eqref{xi-2}, is the same accidental symmetry originally found in Ref.~\cite{Porfyriadis:2021psx} because, as we will now show, it simply acts on the same solution $h$ in Eq.~\eqref{h100}, albeit written in a different gauge, and produces the solution $H$ in Eq.~\eqref{H naive} also written in a different, correspondingly aligned, gauge.

As mentioned earlier, the most general shift of gauge,
\begin{align}
	h\to h+\mathcal{L}_\zeta g^{\rm BR},
\end{align} 
may be achieved by starting with the one-parameter family of coordinates in Eq.~\eqref{ERN general coord map}. Following this one up by the transformation in Eq.~\eqref{xi diffeo}, which is to say writing RN in the one-parameter family of coordinates,
\begin{align}
	\begin{aligned}
	&\hat{t}={t\over\lambda}+\zeta^t(t,r)+{\kappa^2\over\lambda}\xi^t(t,r)
	+\kappa^2\left(\xi^t(t,r)\partial_t\zeta^t(t,r)+ \xi^r(t,r)\partial_r\zeta^t(t,r)\right)\\
	&\hat{r}=1+\lambda r+\lambda^2\zeta^r(t,r) +\lambda\kappa^2 \xi^r(t,r) +\lambda^2\kappa^2 \left(\xi^t(t,r)\partial_t\zeta^r(t,r)+ \xi^r(t,r)\partial_r\zeta^r(t,r)\right)
	\end{aligned}
\end{align}
and expanding in $\lambda$ and $\kappa$, we now obtain Eq.~\eqref{RN expansion scheme} with $h$ given by Eq.~\eqref{h100-general} and $G$ still given by Eq.~\eqref{G}, while $H$ now depends on both $\zeta$ and $\xi$. As before, $\mathcal{E}(g^{\rm BR}, h)=\mathcal{E}(g^{\rm BR}, G)=0$ for arbitrary $\zeta$ and $\xi$, and for the accidental symmetry \eqref{xi-2}, which sets $G=0$, we also have that $\mathcal{E}(g^{\rm BR}, H)=0$ for any $\zeta$. In this case, $H$ is the solution in Eq.~\eqref{H naive} written in some other gauge, with the choice of gauge being controlled by $\zeta$.
We now choose an appropriate gauge for $h$ in Eq.~\eqref{h100-general} such that $\mathcal{L}_{\Xi}h$ is a linear solution around BR, $\mathcal{E}(g^{\rm BR},\mathcal{L}_{\Xi} h)=0$.\footnote{For Eqs.~\eqref{h100} and \eqref{xi-2}, $\mathcal{L}_{\Xi}h$ is not a solution, $\mathcal{E}(g^{\rm BR},\mathcal{L}_{\Xi} h)\neq 0$.}
The most general such choice of gauge may be written as:
\begin{align}\label{zeta}
	\begin{aligned}
	&\zeta^t= rt + q_1''(t)+ r^3 q_2'(r)+ r^3\partial_r^2 Z(t,r)+ \partial_t^2 Z(t,r)/r\\
	&\zeta^r= 2r^4\partial_r\left(\partial_t Z(t,r)/r\right)
	\end{aligned}
\end{align}
for arbitrary functions $q_1(t)$, $q_2(r)$, and $Z(t,r)$. This choice of $\zeta$ produces the solution $H$ in a gauge that is not properly aligned with  $\mathcal{L}_{\Xi}h$. 
However, the two solutions are the same up to a shift of gauge. As a result, we may set $H=\mathcal{L}_{\Xi}h$ by adjusting the gauge for $H\to H+\mathcal{L}_\chi g^{\rm BR}$ with an appropriate choice of $\chi$ that aligns with the above choice of $\zeta$. Notice that we may ensure that $H$ appears in the double expansion in Eq.~\eqref{RN expansion scheme} in the most general gauge parameterized by $\chi$ by replacing, e.g., Eq.~\eqref{xi diffeo} with  
\be
\begin{aligned} 
t&\to t+\kappa^2\xi^t(t,r)+\lambda\kappa^2 \chi^t(t,r)\\
r &\to r+\kappa^2\xi^r(t,r)+\lambda\kappa^2 \chi^r(t,r).
\end{aligned}
\ee
We find that the appropriate choice of $\chi$ that aligns the gauge for $H$ with the gauge of $h$ associated with Eq.~\eqref{zeta}, and therefore sets $H=\mathcal{L}_{\Xi}h$, is given by
\begin{align}\label{chi}
	\begin{aligned}
	&\chi^t= -{t\over r}+ 2q_1(t) + {q_1''(t)\over r^2} +2q_2(r)+ r q_2'(r)+ 2\partial_r Z(t,r)+ r\partial_r^2 Z(t,r)+ {\partial_t^2 Z(t,r)\over r^3}\\
	&\chi^r=1-2r q_1'(t)-2r\partial_t\partial_rZ(t,r).
	\end{aligned}
\end{align}
For example, a simple pair of extreme anabasis perturbation $h$ and near-extreme anabasis perturbation $h+\kappa^2 H$, which are gauge adjusted versions of Eqs.~\eqref{h100} and \eqref{H naive}, respectively, and which are related by the accidental symmetry $\Xi$ in Eq.~\eqref{xi-2} via $H=\mathcal{L}_{\Xi}h$, may be obtained by setting $q_1(t)=q_2(r)=Z(t,r)=0$ so that $\zeta=rt\,\partial_t$, $\chi=-(t/r)\partial_t+\partial_r$, and
\begin{align}
\begin{aligned}
	h_{\mu\nu}{\rm d}x^\mu {\rm d}x^\nu&=-2r^2 t \, {\rm d}t {\rm d}r+\frac{2}{r}{\rm d}r^2+2r \, {\rm d}\Omega^2 \label{example} \\
	H_{\mu\nu}{\rm d}x^\mu {\rm d}x^\nu&=r^3 t^2 {\rm d}t^2 +\frac{1}{6}r^2\left(5t^3-\frac{21}{r^2}\right){\rm d}t {\rm d}r +\frac{1}{2r}\left(t^2-\frac{3}{r^2}\right){\rm d}r^2 -\frac{1}{2}r\left(t^2-\frac{1}{r^2}\right) {\rm d}\Omega^2.
\end{aligned}
\end{align}

\subsection{Virasoro algebra}

The precise form that the generator of the accidental symmetry takes, relating extreme to near-extreme anabasis perturbations, depends on the choice of gauge one begins with when writing the extreme solution. The choice in Eq.~\eqref{example} corresponds to the generator in Eq.~\eqref{xi-2}. On the other hand, as mentioned previously, writing the extreme anabasis perturbation in Fefferman-Graham gauge yields a vector field generator for the same accidental symmetry of the form $\xi\propto (t^3+9t/r^2)\partial_t-3r(t^2-1/r^2)\partial_r$ \cite{Porfyriadis:2021psx}. It is not difficult to see that, following a direct search as in Ref.~\cite{Porfyriadis:2021psx}, one can find the accidental symmetry $\xi$  for any choice of gauge $\zeta$ in Eq.~\eqref{h100-general}. 

In this paper we singled out the gauge choices that correspond to the accidental symmetry in the form of the vector field $\xi_{-2}$ in Eq.~\eqref{xi-2}. In this form, together with the ${\rm SL}(2)$ generators in Eq.~\eqref{SL2 algebra}, we have all the symmetries of the linearized Einstein equation around BR embedded into the Virasoro generating vector fields,
\begin{align}\label{Virasoro generators}
	\xi= \left(\epsilon(t)+\frac{\epsilon''(t)}{2r^{2}}\right)\partial_{t}-\left(r\epsilon'(t)-\frac{\epsilon'''(t)}{2r}\right)\partial_{r}.
\end{align}
Indeed, the modes $\epsilon(t)=t^{-n+1}$ satisfy the Virasoro algebra,
\begin{align}\label{Virasoro algebra}
	[\xi_m,\xi_n] = (m-n)\xi_{m+n},
\end{align}
with $\xi_0$ and $\xi_{\pm 1}$ corresponding to the ${\rm SL}(2)$ isometries and $\xi_{-2}$ to the accidental symmetry, respectively. The generators \eqref{Virasoro generators}, looked at from the point of view of an expansion in $1/r$, have all the expected features of asymptotic symmetry group generators for ${\rm AdS}_2$~\cite{Hotta:1998iq, Cadoni:1999ja, Mandal:2017thl}: they enhance its ${\rm SL}(2)$ isometry group and act as time reparametrizations $t\to t+\epsilon(t)$ on its boundary.\footnote{In Ref.~\cite{Porfyriadis:2021psx}, an attempt was made to write an analogous Virasoro generator for accidental symmetries acting in Fefferman-Graham gauge. However, this attempt had the undesirable feature that it involved nonlinearities in $\epsilon(t)$.} Hence, accidental symmetries of the linearized Einstein equation around ${\rm AdS}_2$ may be thought of as solutions to the equation constraining the allowed asymptotic symmetries, $\epsilon''''=0$. As explained in Ref.~\cite{Porfyriadis:2021psx}, this constraint may also be derived as an equation of motion from a Schwarzian action in the context of Jackiw-Teitelboim (JT) gravity. As a result, accidental symmetries of the Einstein equation around ${\rm AdS}_2$ are also usefully thought of as on-shell large diffeomorphisms (asymptotic symmetries).

\section{Accidental symmetry in the near-horizon of extreme Kerr}\label{sec:Kerr}

We now turn to the considerably more complicated---but more astrophysically relevant---case of spinning black holes.
The near-horizon of extreme Kerr (NHEK) is an ${\rm AdS}_2$-like spacetime that is also an exact solution of the vacuum Einstein equation on its own. The isometry group of NHEK is ${\rm SL}(2)\times {\rm U}(1)$, generated by the Killing vectors
\begin{align}\label{NHEK Killing}
	\xi_{1} = \partial_t,\quad
	\xi_0 = t \partial_t - r \partial_r ,\quad
	\xi_{-1} = \left(t^2 + \frac{1}{r^2}\right)\partial_t - 2 r t \partial_r - \frac{2}{r}\partial_\phi,\quad
	\eta_{(\phi)}=\partial_\phi,
\end{align}
with $\xi_0\,, \xi_{\pm 1}$ obeying the ${\rm SL}(2)$ algebra \eqref{SL2 algebra} and $\eta_{(\phi)}$ corresponding to the rotational ${\rm U}(1)$.
Similar to the BR case, linear axisymmetric gravitational perturbations of NHEK include extreme and near-extreme anabasis solutions as well, corresponding to back reacting NHEK towards building extreme or near-extreme Kerr, respectively. As a result, one expects that there exists an accidental symmetry of the linear Einstein equation around NHEK that enhances its isometry group and allows for mapping extreme to near-extreme linear solutions.
In this section, we find this symmetry for NHEK following the method devised in the previous section for BR.

Consider the Kerr black hole solution in Boyer-Lindquist coordinates
\begin{align}
	{\rm d}s^2=-{\Delta \over\hat \rho^2}\left({\rm d}\hat t-a \sin^2\theta\, {\rm d}\hat\phi\right)^2 +{\sin^2 \theta \over \hat \rho^2}
	\left((\hat r^2+a^2){\rm d}\hat \phi-a \,{\rm d}\hat t\right)^2+{\hat\rho^2 \over\Delta}{\rm d}\hat r^2+\hat \rho^2 {\rm d}\theta^2,
\end{align}
with $\Delta=\hat r^2-2M\hat{r}+a^2$ and $\hat \rho^2=\hat r^2 +a^2\cos^2 \theta$. Extreme Kerr (EK) is defined by $a=M$, corresponding to the maximal allowed angular momentum $J=aM\leq M^2$ for a given mass.
For EK, one may use the following one-parameter family of coordinates, in $M=1$ units,
\begin{align}\label{EK coord map}
	\hat{t}=2t/\lambda, \quad \hat{r}=1+\lambda r, \quad \hat{\phi}=\phi+t/\lambda,
\end{align}
in order to obtain the series
\begin{align}\label{EK series}
	g^{EK}=g^{\rm NHEK}+\lambda \, h +\mathcal{O}(\lambda^2),
\end{align}
with the NHEK metric obtained in Poincar\'e coordinates
\be 
\begin{aligned}\label{NHEK Poincare}
	g^{\rm NHEK}_{\mu\nu}{\rm d}x^\mu {\rm d}x^\nu&= 
	\gamma(\theta)\left( -r^2 {\rm d}t^2 + {{\rm d}r^2 \over r^2} + {\rm d}\theta^2 \right) + \nu(\theta)({\rm d}\phi + r\, {\rm d}t)^2,\\
	&\gamma(\theta)=1+\cos^2\theta, \quad \nu(\theta)=\frac{4\sin^2\theta}{1+\cos^2\theta}, 
\end{aligned}
\ee
and $h$ given by
\begin{align}\label{h100 NHEK}
	\begin{aligned}
		h_{\mu\nu}{\rm d}x^\mu {\rm d}x^\nu =
		&- \left(\frac{\nu^2}{8}-\frac{8(\gamma-1)}{\gamma^2}\right) r^3 {\rm d}t^2 
		+ \frac{2}{r} {\rm d}r^2 + 2 r\,{\rm d}\theta^2 
		\\
		&+ \nu \left(1+\frac{\gamma-\nu}{2}\right) r^2 {\rm d}t {\rm d}\phi
		+ 2 \nu \frac{\gamma-1}{\gamma} r \, {\rm d} \phi^2.
	\end{aligned}
\end{align}
Clearly, as a solution of the linearized Einstein equation around NHEK, $\mathcal{E}(g^{\rm NHEK},h)=0$, the perturbation $h$ is an extreme anabasis perturbation whose back reaction on NHEK builds EK.

Now consider a near-extreme Kerr with
\begin{align}
	{a\over M}=\sqrt{1-\lambda^2\kappa^2}.
\end{align}
Following up the transformation \eqref{EK coord map} by
\begin{align}\label{xi diffeo NHEK}
	t\to t+\kappa^2\xi^t(t,r), \quad 
	r\to r+\kappa^2\xi^r(t,r), \quad
	\phi\to \phi+\kappa^2\xi^\phi(t,r),
\end{align}
and using the resulting one-parameter family of coordinates 
\begin{align}\label{Kerr nice coords}
	\begin{aligned}
		&\hat{t}={2t\over\lambda}+{2\kappa^2\over\lambda}\xi^t(t,r)\\ &\hat{r}=1+\lambda r+\lambda\kappa^2 \xi^r(t,r)\\
		&\hat{\phi}=\phi+\kappa^2\xi^\phi(t,r)+{t\over\lambda}+	{\kappa^2\over\lambda}\xi^t(t,r),
	\end{aligned}
\end{align}
one can write the Kerr solution as the double series in $\lambda$ and $\kappa$,
\begin{align}\label{Kerr expansion scheme}
	g^{\rm Kerr}=g^{\rm NHEK}+\kappa^2\,G +\lambda h +\lambda\kappa^2\,H+\cdots.
\end{align}
Here, $g^{\rm NHEK}$ and $h$ are given as in Eqs.~\eqref{NHEK Poincare} and \eqref{h100 NHEK}, respectively, while $G$ is given by
\begin{align} \label{G Kerr}
	\begin{aligned}
		G_{tt}&=
		\gamma-2(\gamma-\nu) \,r (\xi^r+r\partial_t\xi^t)+2\nu r \partial_{t}\xi^\phi  \\
		G_{tr}&=-(\gamma-\nu)r^2 \partial_r\xi^t 
		+\gamma\frac{1}{r^2}\partial_t\xi^r 
		+\nu \, r\partial_{r}\xi^\phi \\
		G_{rr}&=\gamma\left(\frac{1}{r^4}-\frac{2}{r^3}\left(\xi^r-r\partial_r\xi^r\right)\right)\\
		G_{t\phi}&=\nu(\xi^r+r\partial_t\xi^t+\partial_t\xi^\phi) \\		
		G_{r\phi}&=\nu(r\partial_r\xi^t+\partial_r\xi^\phi).
	\end{aligned}
\end{align}
As in the RN case, we have that $G$ solves the linearized Einstein equation around NHEK, $\mathcal{E}(g^{\rm NHEK},G)=0$, for arbitrary $\xi$; if we want to have $H$ in Eq.~\eqref{Kerr expansion scheme}, which also depends on $\xi$, to be a solution too, then we need to set $G=0$. Doing so we find
\begin{align}\label{xi-2 Kerr with SL2xU1 added}
	\xi=\frac{1}{12}\left(t^3+\frac{3t}{r^2}\right)\partial_t
	-\frac{1}{4}r\left(t^2-\frac{1}{r^2}\right)\partial_r
	-\frac{t}{2r}\partial_\phi
	+e_0\xi_0+e_1\xi_{1}+e_{-1}\xi_{-1}+e_{(\phi)}\eta_{(\phi)}.
\end{align}
Hence, we see that, in addition to the ${\rm SL}(2)\times {\rm U}(1)$ isometries of NHEK, we have the accidental symmetry,
\begin{align}\label{xi-2 Kerr}
	\Xi=\frac{1}{12}\,\xi_{-2}  \quad\textrm{with}\quad
	\xi_{-2}=\left(t^3+\frac{3t}{r^2}\right)\partial_t
	-3r\left(t^2-\frac{1}{r^2}\right)\partial_r
	-\frac{6t}{r}\partial_\phi.
\end{align}
Using this accidental symmetry, the expansion in Eq.~\eqref{Kerr expansion scheme} becomes
\begin{align}\label{Kerr series naive gauge}
	g^{Kerr}=g^{\rm NHEK}+\lambda \left(h +\kappa^2\,H\right)+\cdots
\end{align}
with
\begin{align}\label{H naive Kerr}
	\begin{aligned}
		H_{\mu\nu}{\rm d}x^\mu {\rm d}x^\nu=& 
		\left[\left(\frac{1}{2}-\frac{4(\gamma-1)}{\gamma^2}\right)r^3\left(t^2-\frac{1}{r^2}\right)-\frac{\nu(\gamma+2)}{4}r\right]{\rm d}t^2
		-\frac{1}{2r}\left(t^2-\frac{1}{r^2}\right){\rm d}r^2
		\\
		&-\left(\gamma+\frac{4(\gamma-1)}{\gamma}\right)t\, {\rm d}t{\rm d}r 
		-\frac{1}{2}r\left(t^2-\frac{1}{r^2}\right){\rm d}\theta^2
		-\frac{\nu(\gamma-1)}{2\gamma}r\left(t^2-\frac{1}{r^2}\right){\rm d}\phi^2
		\\
		&-\frac{\nu}{4}\left[\gamma+2+\left(1+\frac{\gamma-\nu}{2}\right)r^2\left(t^2-\frac{1}{r^2}\right)\right] {\rm d}t{\rm d}\phi
		+\frac{\gamma\nu^2}{16}\frac{t}{r}\, {\rm d}r{\rm d}\phi.
	\end{aligned}
\end{align}
From Eq.~\eqref{Kerr series naive gauge}, it is clear that $h+\kappa^2 H$, considered as a solution of the linearized Einstein equation around NHEK, $\mathcal{E}(g^{\rm NHEK},h+\kappa^2 H)=0$, is a near-extreme anabasis perturbation off NHEK whose back reaction builds a near-extreme Kerr geometry.

The NHEK accidental symmetry $\xi_{-2}$ in Eq.~\eqref{xi-2 Kerr}, together with its  ${\rm SL}(2)$ generating Killing vectors $\xi_0, \xi_{\pm 1}$ in Eq.~\eqref{NHEK Killing}, may be embedded into the Virasoro generating vectors fields,
\begin{align}\label{Virasoro generators NHEK}
	\xi= 
	\left(\epsilon(t)+\frac{\epsilon''(t)}{2r^{2}}\right)\partial_{t}
	-\left(r\epsilon'(t)-\frac{\epsilon'''(t)}{2r}\right)\partial_{r}
	-\frac{\epsilon''(t)}{r}\partial_\phi\,,
\end{align}
using the modes $\epsilon(t)=t^{-n+1}$. These obey the Virasoro algebra \eqref{Virasoro algebra} and have previously been proposed, in the context of the Kerr/CFT correspondence \cite{Guica:2008mu, Compere:2012jk}, as potential asymptotic symmetry group generators for NHEK\cite{Matsuo:2009sj, Matsuo:2010ut}.

\subsection{Moving to a singular gauge}

The accidental symmetry $\Xi$, given by Eq.~\eqref{xi-2 Kerr}, does not relate the extreme and near-extreme anabasis perturbations $h$ and $h+\kappa^2 H$, given by Eqs.~\eqref{h100 NHEK} and \eqref{H naive Kerr}, via $H=\mathcal{L}_{\Xi}h$.\footnote{Indeed, for Eqs.~\eqref{h100 NHEK} and \eqref{xi-2 Kerr}, $\mathcal{L}_{\Xi}h$ is not even a solution, $\mathcal{E}(g^{\rm NHEK},\mathcal{L}_{\Xi} h)\neq 0$.}  
However, as in the RN case, it does relate gauge adjusted versions of them. For example, shifting the gauge of Eq.~\eqref{h100 NHEK} by $\mathcal{L}_{\zeta}g^{\rm NHEK}$ with 
\begin{align}\label{zeta for NHEK example}
	\zeta=\left(-\frac{1}{4}r^2\cos^2\theta-r^2\log{r}\right)\partial_{r}
	+\frac{r (217+56\cos{2\theta}-\cos{4\theta})}{16 \sin{2\theta}}\partial_\theta,
\end{align}
we obtain
\begin{align}\label{NHEK example h}
\begin{aligned}
	h_{\mu\nu}{\rm d}x^\mu {\rm d}x^\nu=
	\frac{r^3\left(3(129+\log{r})+4(10+7\log{r})\cos{2\theta}+(5+\log{r})\cos{4\theta}\right)}{2(3+\cos{2\theta})} \, {\rm d}t^2
	\\
	+\frac{2r^2\left(147-4\log{r}-(11-4\log{r})\cos{2\theta}\right)}{3+\cos{2\theta}} \, {\rm d}t {\rm d}\phi
	+\frac{4r(37-3\cos{2\theta})}{3+\cos{2\theta}} {\rm d}\phi^2
	\\
	-\frac{1}{r}(3+\cos{2\theta})(5+\log{r}) \, {\rm d}r^2
	+\frac{(3+\cos{2\theta})^2(37-3\cos{2\theta})}{8\sin{2\theta}}	{\rm d}r {\rm d}\theta
	\\
	+\frac{1}{16}r\left(249-44\cos{2\theta}+3\cos{4\theta}-544\csc^2\theta+160\sec^2\theta\right) {\rm d}\theta^2,
\end{aligned}
\end{align}
for which it is easy to verify that $\mathcal{L}_{\Xi}h$ is nothing other than a gauge adjusted version of the solution in Eq.~\eqref{H naive Kerr}.

Notice that the diffeomorphism generated by Eq.~\eqref{zeta for NHEK example} is singular on the sphere. We have searched for the most general diffeomorphism that may be used to adjust the gauge for the regular solution in Eq.~\eqref{h100 NHEK} in such a way that in this gauge $\mathcal{L}_{\Xi}h$ is a solution of the linearized Einstein equation around NHEK; that is to say, we have searched for all possible $\zeta$ such that $\mathcal{E}(g^{\rm NHEK},\mathcal{L}_{\Xi} (h+\mathcal{L}_{\zeta}g^{\rm NHEK}))=0$ with $h$ given by Eq.~\eqref{h100 NHEK} and $\Xi$ the accidental symmetry in Eq.~\eqref{xi-2 Kerr}.
The result may be written as follows:
\begin{align}\label{general zeta}
\begin{aligned}
	\zeta^t =& -\frac{1}{32r}\left(\ddot T(t,r) + r^4 T''(t,r)\right) \\
	\zeta^r =&
	-\frac{r^2(1+\cos2\theta+8\log r)}{8} 
	-\frac{r}{32}\left(\dddot L(t,r)+r^4 \dot L''(t,r)\right)
	\\
	& +r^2 \beta_0 \left(\cos\theta+2\cot\theta\csc\theta-2\,{\rm arctanh}(\cos\theta)\right) + r^2 \beta_1 \\
	\zeta^\theta=&\,
	\frac{r\csc 2\theta}{16}(217+56\cos 2\theta-\cos 4\theta)
	-\frac{(3+\cos 2\theta)\csc 2\theta}{32}\left(\dddot L(t,r)-r^4 \dot L''(t,r)\right)\\
	&-(4r\beta_0-\beta_2)(3+\cos2\theta)\csc 2\theta\, {\rm arctanh}(\cos\theta) + r\beta_0
	(4\csc^4 \theta-1)\sin\theta \\
	& +\frac{1}{8}\beta_3 (2\cot\theta+\tan\theta)+r(3+\cos2\theta)\csc 2\theta \left(W_+\left(it+{1}/{r}\right) + W_-\left(it-{1}/{r}\right)\right).
\end{aligned}
\end{align}
Here an overdot and a prime denote  differentiation with respect to $t$ and $r$, respectively, and $T(t,r) = r^3 \partial_r(L(t,r)/r)$, while the three functions $L, W_\pm$ and the four constants $\beta_{0,1,2,3}$ are arbitrary. It is interesting that, depending on how we fix the arbitrary parameters in Eq.~\eqref{general zeta}, we can choose the coordinate singularity to be localized on the equator or the poles but we can never remove it completely. In other words, in order to have the relation $H=\mathcal{L}_{\Xi}h$, we necessarily need to move the regular solutions in Eqs.~\eqref{h100 NHEK} and \eqref{H naive Kerr} to a singular gauge. 

We emphasize that we are considering the excursion to singular gauges generated by Eq.~\eqref{general zeta} only for the purpose of having the accidental symmetry $\Xi$, found in Eq.~\eqref{xi-2 Kerr}, act via the Lie derivative on an extreme anabasis perturbation and generate explicitly a near-extreme one. Such action on the perturbative solutions of the near-horizon of near-extreme black holes was taken in Ref.~\cite{Porfyriadis:2021psx} to be the definition of an accidental symmetry. However, in light of the enhanced understanding presented in this paper, we can equivalently define accidental symmetries to be simply any nontrivial solutions of the equations $G=0$ in an expansion such as Eq.~\eqref{RN expansion scheme} for RN or Eq.~\eqref{Kerr expansion scheme} for Kerr.

A similar excursion to singular gauges was taken in Ref.~\cite{Castro:2021csm} when attempting to mold the gravitational perturbations of NHEK into the form of JT equations for nearly-${\rm AdS}_2$ holography. Indeed, Ref.~\cite{Castro:2021csm} found it worth writing regular anabasis perturbations, such as the ones in Eqs.~\eqref{h100 NHEK} and \eqref{H naive Kerr}, as a sum of a singular perturbation $\mathcal{L}_{\xi}g^{\rm NHEK}$, generated by a non-single valued diffeomorphism $\xi$, with another singular perturbation of NHEK in such a way that both may be identified with a JT mode.
We leave further investigation of this particular analogy as well as the significance of excursions to singular gauges in general to future work.

\section{Discussion}\label{sec:discussion}

In this paper, we have found a new, accidental symmetry of the vacuum Einstein equation for perturbations of the NHEK geometry describing the near-horizon region of an extremal Kerr black hole.
To do so, we first reconsidered the RN case treated in Ref.~\cite{Porfyriadis:2021psx}, discovering a novel method of mechanically deriving the accidental symmetry by means of a scaling coordinate transformation from RN to the near-horizon BR space that---crucially unlike the usual Rindler versus Poincar\'e  ${\rm AdS}_2$ coordinates for finite versus vanishing temperature---is regular in the zero-temperature limit.
We showed how such a regular coordinate transformation leads to an accidental symmetry that acts on the anabasis perturbation off of BR, in a way that transforms zero-temperature solutions to finite-temperature ones.
We then treated the analogous question for Kerr black holes, discovering a new accidental symmetry of the NHEK geometry that connects the extremal anabasis to near-extremal solutions.
Remarkably, the accidental symmetry beautifully unifies with the ${\rm SL}(2)$ Killing vectors of the near-horizon ${\rm AdS}_2$ into a Virasoro algebra.

This work suggests several avenues for future investigation.
First, it is notable that the NHEK accidental symmetry obtained in this way can only be made to act explicitly in a gauge with a coordinate singularity either at the poles or equator of the black hole, and it would be interesting to ascertain why this must be the case.
Further, while the fact that the output of the action of the accidental symmetry is necessarily a finite-temperature solution is clear---since we match this solution itself to one obtained from a scaling limit of an explicit finite-temperature Kerr geometry---it would be useful to identify a gauge invariant observable that can identify the temperature purely in the language of NHEK perturbations alone, analogous to the invariant ${\rm SL}(2)$ Casimir $\mu$ in the RN case.
We leave this question, along with the related study of the behavior of the corresponding Weyl scalars in the Newman-Penrose formalism, to future work.
Additionally, in Ref.~\cite{Porfyriadis:2021psx} it was found, remarkably, that dynamical perturbative wave solutions of a minimally coupled scalar in the BR background could themselves be described as another type of accidental symmetry acting on the extremal anabasis solution.
The challenging question of whether such an understanding of gravitational waves on the NHEK background can be achieved in terms of more exotic accidental symmetries---and whether this could be useful for calculations of astrophysical interest---is another direction that we will investigate in future work.

Moreover, when considering the near-horizon limit of finite-energy dynamical perturbations of asymptotically flat extreme black holes, it is often the case that part of the near-horizon approximation of these solutions takes the same form as the near-extreme anabasis perturbations off the ${\rm AdS}_2$ throat geometry of the background, making it difficult to disentangle the dynamical from the anabasis part of the resulting perturbative ${\rm AdS}_2$ solution~\cite{deCesare:2024csp}. It would therefore be interesting to use our improved understanding of the role that accidental symmetries play in mapping perturbative ${\rm AdS}_2$ solutions among themselves, in order to manifest such disentangling.
Finally, in light of the high curvature and ultraviolet sensitivity recently discovered for extremal horizons in gravitational effective theory~\cite{Horowitz:2023xyl,Horowitz:2024dch}, it would also be of interest to understand how accidental symmetries behave under the addition of higher-derivative corrections.

\bigskip

\begin{center} 
{\bf Acknowledgments}
\end{center}
\noindent 
We thank Shahar Hadar, Dan Kapec, and Alex Lupsasca for useful discussions and comments.
This research is implemented in the framework of H.F.R.I. call ``Basic research Financing (Horizontal support of all Sciences)'' under the National Recovery and Resilience Plan ``Greece 2.0'' funded by the European Union --- NextGenerationEU (H.F.R.I. Project Number: 15384). APP is also partially supported by the European MSCA grant HORIZON-MSCA-2022-PF-01-01.
G.N.R. is supported by the James Arthur Postdoctoral Fellowship at New York University.

\vspace{3mm}

\bibliographystyle{utphys-modified}
\bibliography{NHEKsym}

\end{document}